\documentclass[pra,twocolumn,hidelinks,superscriptaddress,byrevtex,titlepage,secnumarabic,nofootinbib,longbibliography,a4paper,showkeys,preprintnumbers]{revtex4-2}
\usepackage{graphicx}
\usepackage{amsmath}
\usepackage{amssymb}
\usepackage{float}
\usepackage{moreverb}
\usepackage{textcomp}
\usepackage{url}
\usepackage{scrextend}
\usepackage{xcolor}
\usepackage{verbatim}
\usepackage{colortbl}
\usepackage{mhchem}
\usepackage[datesep=,datetimesep=.,timesep=,showisoZ=true,showseconds=false]{datetime2}
\usepackage[pdftex,
            pdfauthor={Jose Maria Martin-Olalla et al},
            pdftitle={Ultraslow calorimetric studies of the martensitic transformation of NiFeGa alloys: detection and analysis of avalanche phenomena},
            pdfsubject={martensitic transformations},
            citecolor=.,
            pdfkeywords={},
            runcolor=black,
            linkcolor=blue,
            pdfproducer={Latex with hyperref, or other system},
            pdfcreator={pdflatex, or other tool}]{hyperref}

            \hypersetup{colorlinks,
              linkcolor={blue}}
\usepackage{breakurl}
\usepackage{siunitx}
\DeclareSIUnit\permille{\text{\textperthousand}}
\usepackage{mathtools}
\usepackage[utf8]{inputenc}
\usepackage{lineno}
\usepackage{array}
\AtBeginDocument{\usepackage{booktabs}}

\setcitestyle{numbers,square}
\makeatletter\AtBeginDocument{\let\@elt\relax}\makeatother

\newcommand{\usample}{\ce{Ni_{55}Fe_{19}Ga_{26}}}

\newcommand{\ufcomment}[1]{}
\newcommand{\uaddress}{Departamento de Física de la Materia Condensada, Facultad de Física, ICMSE CSIC, Universidad de Sevilla,  ES41012 Sevilla, Spain}

\newcommand{\ualtaddress}{Instituto de Ciencia de Materiales de Sevilla, ICMSE CSIC, Universidad de Sevilla, Sevilla, Spain}
\begin{document}
\author{José-María Martín-Olalla}
\email{olalla@us.es}
\homepage{https://orcid.org/0000-0002-3750-9113}
\homepage{https://ror.org/03yxnpp24}
\affiliation{\uaddress}
\author{Antonio Vidal-Crespo}
\affiliation{\uaddress}
\author{Francisco Javier Romero}
\affiliation{\uaddress}
\author{Alejandro F. Manchón-Gordón}
\affiliation{\ualtaddress}
\author{Jhon J. Ipus}
\affiliation{\uaddress}

\author{Javier S. Blázquez}
\affiliation{\uaddress}
\author{María Carmen Gallardo}
\affiliation{\uaddress}
\author{Clara F. Conde}
\affiliation{\uaddress}
\date[Sent: ]{4 October 2023; Revised: 15 March 2024; Accepted: 8 April 2024; Publised: 19 May 2024}
\title{Ultraslow calorimetric studies of the martensitic transformation of NiFeGa alloys: detection and analysis of avalanche phenomena}

\preprint{This is the accepted version and does not reflect post-acceptance improvements, or any corrections. \copyright The Authors }
\preprint{The Version of Record (OA) is published  in the \emph{Journal of Thermal Analysis and Calorimetry} \doi{10.1007/s10973-024-13206-4}.}


\newcommand{\NTms}{327}
\newcommand{\NTmf}{315}
\newcommand{\NTas}{322}
\newcommand{\NTaf}{335}
\newcommand{\NTdhm}{-2.7}
\newcommand{\NTdha}{-2.9}
\newcommand{\NTpeakd}{322.7}
\newcommand{\NTpeakr}{329.8}
\newcommand{\hNT}{7.1}
\newcommand{\Tms}{317.3}
\newcommand{\Tmf}{308.4}
\newcommand{\Tas}{314.2}
\newcommand{\Taf}{323.6}
\newcommand{\Tdhm}{-3.1}
\newcommand{\Tdha}{-3.3}
\newcommand{\Tpeakd}{313.1}
\newcommand{\Tpeakr}{319.6}
\newcommand{\hT}{6.5}

\begin{abstract}
  We study the thermal properties of a bulk \usample{} Heusler alloy in a conduction calorimeter. At slow heating and cooling rates ($\sim\SI{1}{\kelvin\per\hour}$), we compare as-cast and annealed samples. We report a smaller thermal hysteresis after the thermal treatment due to the stabilization of the 14M modulated structure in the martensite phase. In ultraslow experiments (\SI{40}{\milli\kelvin\per\hour}), we detect and analyze the calorimetric avalanches associated with the direct and reverse martensitic transformation from cubic to 14M phase. This reveals a distribution of events characterized by a power law with exponential cutoff $p(u)\propto u^{-\varepsilon}\exp(-u/\xi)$ where $\varepsilon\sim 2$ and damping energies $\xi=\SI{370}{\micro\joule}$ (direct) and $\xi=\SI{27}{\micro\joule}$ (reverse) \textcolor{black}{that characterize the asymmetry of the transformation}.

\begin{verbatim}
File: mainAps.tex
Encoding: utf8
Words in text: 4403
Words in headers: 34
Words outside text (captions, etc.): 758
Number of headers: 12
Number of floats/tables/figures: 10
Number of math inlines: 177
Number of math displayed: 2
Subcounts:
  text+headers+captions (#headers/#floats/#inlines/#displayed)
  155+17+0 (2/0/5/0) _top_
  724+1+0 (1/0/12/0) Section: Introduction
  530+1+0 (1/0/8/0) Section: Experimental
  0+1+0 (1/0/0/0) Section: Results
  441+4+157 (1/3/13/0) Subsection: Structural and microstructural characterization
  368+2+177 (1/2/22/0) Subsection: DTA traces
  715+1+235 (1/3/32/0) Subsection: Avalanches
  1034+1+189 (1/2/83/2) Section: Discussion
  300+1+0 (1/0/2/0) Section: Conclusions
  52+1+0 (1/0/0/0) Section: Declarations
  84+4+0 (1/0/0/0) Section: CRediT authorship contribution statement


\end{verbatim}
\end{abstract}




\maketitle

    \section{Introduction}
\label{sec:introduction}


\textcolor{black}{\ce{TiNi}-based alloys are extensively favored in various applications owing to their distinctive shape memory effect and superelasticity~\cite{Zhang2006,Fan2004,Thamburaja2002}. Nevertheless, these alloys come with drawbacks such as high cost and a challenging fabrication process~\cite{Mohd2018}. In lieu of these alloys, there has been significant research on \ce{Cu}-based shape memory alloys due to their cost-effectiveness and comparatively straightforward processing~\cite{Canbay2013,Silva2022,Pedrosa2022}. In addition to the conventional thermally induced shape memory effect observed in \ce{TiNi}-based alloys, in 1996,}~\citeauthor{Ullakko1996}~\cite{Ullakko1996} reported a large magnetic field induced strain in \ce{Ni2MnGa} single crystals.  These possible applications are related to the martensitic transformation (MT) that takes place in this kind of materials, a first-order phase transition which occurs in the solid state from a high temperature (high symmetry) austenite phase to a low temperature (low symmetry) martensite phase.

The structure of the martensite phase and the martensitic transformation temperature have been extensively analyzed in \ce{NiMn}-based Heusler alloys~\cite{Planes2007,Bachaga2019}, which are very sensitive to both the chemical composition tailoring~\cite{Maziarz2017} and the fabrication method~\cite{Manchon-Gordon2021}. In this sense, the dependence of the valence electron concentration per atom, $e/a$, is of great significance in the development of Heusler alloys. As an example, the transition temperatures linearly increase with $e/a$ in \ce{Ni-Mn-X} (\ce{X=In, Sn, Ga}) systems, i.e. decreasing the \ce{X} element concentration~\cite{Moya2006}. On the other hand, it has been shown that the crystal structure of the martensite phase evolves in the sequence $10\mathrm{M}\to14\mathrm{M}\to\mathrm{L}1_0$ with the increase of  $e/a$  in \ce{Ni-Mn-In} system~\cite{Moya2006}.

Although the general formula of Heusler alloys includes \ce{Ni, Mn} and one element of the \ce{Ga, In, Sn} or \ce{Sb} quartet, it has been proposed to replace \ce{Mn} by \ce{Fe} in order to improve the mechanical properties~\cite{Pons2008}. The enhanced ductility of these \ce{Ni-Fe-Ga} alloys are related to the precipitation of the secondary $\gamma-$phase. In contrast to the compositional series with \ce{Mn}, the dependence of the transition temperature is not so obvious. For $e/a>7.8$, transition temperature decreases with the increase of $e/a$~\cite{Yu2009,Sarkar2014}. However, for $e/a<7.8$ there is not a clear tendency. Moreover, it has been found that at $e/a=7.8$, the structural transition temperature and both Curie temperatures of the austenite and the martensite phases are close to room temperature~\cite{Manchon-Gordon2021}, which could make this system candidate for room temperature applications. Therefore, it is of the most interest to perform precise analyses of the martensitic transformation of this system.

Low temperature martensitic transformations are characterized by a very fast diffusionless growth \textcolor{black}{that can be detected by magnetic measurements~\cite{Recarte2012,Romero2021a}, thermal measurements~\cite{Bouabdallah2002,Cesari2005,Gallardo2010}, XRD-\emph{in situ} techniques~\cite{Czaja2017}, and electrochemical impedance~\cite{Cunha2019}. The diffusionless growth}  can be described as an autocatalytic kinetic process (e.g. for \ce{NiFeGa}~\cite{Manchon-Gordon2021a} and \ce{NiMnIn} Heusler alloys~\cite{Blazquez2022}) and yields local symmetry changes that are transmitted to the surrounding at the speed of sound. MT are known to occur intermittently~\cite{Vives1994,Planes2013,Salje2014}, due to the existence of kinetic impediments, such as lattice defect, impurities or self-generated heterogeneities that produce a complex energy landscape with different metastable states ---multiple minima separated by high energy barriers--- in the region of coexistence of the high-symmetry and the low-symmetry phases. If the energy barriers are large enough, the transition only happens when the system is externally driven (athermal transition), so that the transition extends over a finite interval temperature. The strain energy can be stored in the lattice elastically and blocks subsequent growth of the new phase. The transition happens by a succession of strain relaxations between metastable states and avalanches are linked to these fast relaxation processes.

The detection of avalanches has traditionally required the use of experimental techniques sensitive to the strain jumps that occur during relaxation such as acoustic emission~\cite{Planes2013}; magnetic Barkhausen emission have also provided results in this field~\cite{Sullivan2004,Baro2013,Toth2016}. Alternatively, calorimetric techniques have proven able to detect and measure the energy associated with the avalanches, which are observed as spikes in DTA traces of high resolution conduction calorimetry~\cite{Gallardo2010,Vives2016,Romero2019,Romero2021a} and in differential scanning calorimetry (DSC) traces~\cite{Blobaum2006,Bolgar2017,Kamel2023}. \textcolor{black}{These techniques provide a straightforward identification of the characteristic energies involved in the jerky events at the prize of more challenging experimental conditions.}

In this work an arc-melted \usample ($e/a=7.8$) Heusler alloy is thermally analyzed before and after a thermal treatment, which leads to the formation of a modulated $14\mathrm{M}$ martensite phase from a previous tetragonal martensite phase $\mathrm{L}1_0$. The \textcolor{black}{characteristic jerky behavior observed in single crystals~\cite{Bolgar2017} is reproduced in the arc-melted polycrystalline sample after the thermal treatment, with events distributed according to a power-law with an exponential cutoff. The characteristic damping energy is smaller (higher damping) in the reverse transformation.}

\section{Experimental}
\label{sec:experimental}

An alloy with nominal, atomic composition \usample{} was prepared from a mixture of high-purity constituent elements ($>\SI{99.9}{\percent}$) in argon atmosphere in an arc furnace MAM-1 (Edmund B\"uhler GmbH). In order to keep the final composition as close as possible to the nominal one, an additional amount of \SI{2}{\percent}    excess of \ce{Ga} was added to compensate losses connected with its evaporation during fabrication process (melting temperature \SI{303}{\kelvin}). \textcolor{black}{To obtain a high chemical homogeneity, the alloy was arc-melted several times inside the furnace}. Then, parts of the obtained ingot were cut and annealed at \SI{1073}{\kelvin} inside a quartz tube under \ce{Ar} pressure, adding some \ce{Zr} wires as getter to prevent oxidation. After an annealing time of \SI{24}{\hour}, the quartz tube was quenched in water.

The chemical composition of the original ingot was analyzed by X-ray fluorescence (XRF) using an EAGLE III instrument with an anticathode of \ce{Rh}.

The average composition of the sample was determined by XRF, obtaining an actual chemical composition (atomic percentage) of \ce{Ni_{54.6}Fe_{19.4}Ga_{26}}. The averaged chemical composition is very close to the nominal one, with a small depletion of \ce{Ni} and a small enrichment of \ce{Fe}. Based on the XRF measurements and on the electronic configuration of the outer shells for each element ---\ce{Ni} (\ce{4s^2 3d^8}), \ce{Fe} (\ce{4s^2 3d^6}), and \ce{Ga} (\ce{4s^2 4p})--- we estimated the electron valence concentration per atom  as $e/a=7.79$.

The crystal structure was investigated by X-ray diffraction (XRD, Bruker D8 I diffractometer) using \ce{Cu} K-alpha radiation. Bruker DIFFRAC.EVA (phase identification) and DIFFRAC.TOPAS (Le Bail refinement) software were employed in order to analyze the XRD patterns. Scanning electron microscopy (SEM) was used to study the microstructural features in a FEI Teneo microscope, using secondary electron (SE), backscattered electron (BSE) modes and energy dispersive X-ray analysis (EDS) to determine the local composition of the different phases. Before the observation in the microscope, the samples were bonded in an epoxy resin and polished.


Thermal properties were studied in a conduction calorimeter described elsewhere~\cite{DelCerro1987b,Jimenez1988}, which can record high resolution DTA traces. For this purpose, the measurement device consists of two fluxmeters, with 48 chromel-constantan thermocouples each, which are disposed electrically in series and thermally in parallel. The signal provided by the fluxmeters $E$ is recorded by a Keithley 182 nanovoltmeter at a sampling rate of $\SI{12.5}{\hertz}$ and is  converted into a heat flux $\phi$ by scaling it with  a sensitivity determined in a calibrating experiment~\cite{Martin1997}. If the rate of temperature change is $\beta$ then the ratio $\phi/\beta$ is the heat exchanged by the sample and the calorimetric block per unit change of temperature. After removing a suitable baseline, any excursion of this quantity is associated with the enthalpy change in the event of phase transition~\cite{Martin-Olalla2000,DelCerro2000,Romero2007}. The system operates then as high resolution differential thermal analyser  with a sensitivity around $\SI{1}{\micro\watt}$. The device is surrounded by  a large thermal bath whose temperature is controlled by a Julabo FP40 or FP45 through a heat exchanger coil. Due to the large thermal inertia of the equipment, the scanning rates range from a few kelvin per hour to few milli kelvin per hour, much slower than commercial differential scanning calorimeters DSC.

\textcolor{black}{The magnetic properties of the \usample{} system were previously studied in Ref.~\cite{Manchon-Gordon2021}. }

\section{Results}
\label{sec:results-1}

\subsection{Structural and microstructural characterization}
\label{sec:struct-micr-char}

\begin{figure*}[t]
  \centering
\includegraphics[scale=0.8]{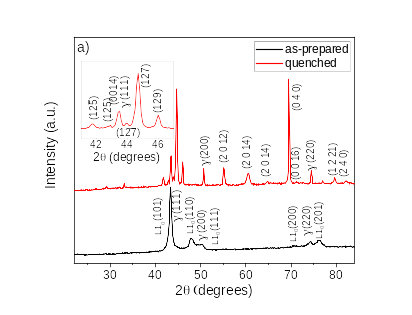}   \includegraphics[scale=0.8]{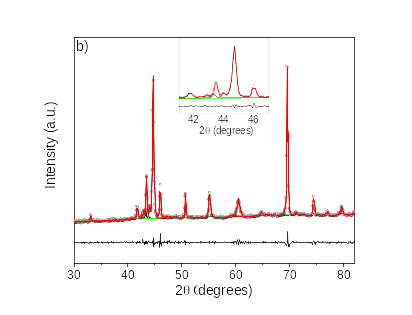}
  \caption{a) XRD patterns of \ce{Ni55Fe19Ga26} as-prepared (ASB) and quenched bulk (TTB) samples. The Bragg peaks of the modulated phase have been labeled according to the notation of the monoclinic system that usually describes the modulated 14M structure. b) XRD pattern of the TTB (open circles) and its Le Bail fitting using  martensite phase and  $\gamma-$phase space groups (red line). The difference is shown at the bottom (black line). The inset depicts a horizontally zoomed plot.}
  \label{fig:xrd}
\end{figure*}

Figure~\ref{fig:xrd} depicts XRD patterns at room temperature of the as-prepared bulk sample (ASB) and after annealing treatment and subsequently quenched in water (Thermally Treated Bulk sample, TTB). It has been found that ASB sample reveals the mixture of a non-modulated martensite L$1_0$ tetragonal structure (space group $I 4/mmm$) with traces of a $\gamma-$phase (space group $Pm\overline{3}m$). This result confirms the limitation of arc-melting technique to produce monophasic compounds. The tetragonal non-modulated martensite, NM, in the ASB sample is replaced by a modulated structure after thermal treatment and subsequent quenching. \emph{Ab initio} calculations suggest that the NM structure is thermodynamically more stable than the 14M structure~\cite{Zayak2004}. Therefore, the thermal stress, excess of vacancies, etc. induced by the quenching process helps to form the non-thermodynamically stable 14M structure at room temperature, which is typically found in Heusler alloys prepared by rapid quenching techniques~\cite{Manchon-Gordon2020}.

\textcolor{black}{In order to evaluate the modulated character of the martensite phase, a Le Bail refinement, which does not require the full crystal structure data but only the space group of the phases, has been performed on the XRD pattern corresponding to the TTB sample (see Figure~ (see Figure~\ref{fig:xrd}b) with GOF=2.03.} The modulated structure can be indexed with the monoclinic  space group, obtaining lattice parameters $a=\SI{0.44553(3)}{\nano\meter}, b=\SI{3.0329(3)}{\nano\meter}, c=\SI{0.56195(3)}{\nano\meter}$ and monoclinic angle equal to $\ang{88.009(7)}$. The relation $b=7a$ indicates a seven fold increase in the unit cell length along  the $b$ axis, coherently with a modulation 14M. We obtained a lattice parameter $a=\SI{0.36053(3)}{\nano\meter}$ for the cubic structure corresponding to the $\gamma-$phase.

\begin{figure*}[t]
  \centering
\includegraphics[scale=0.3]{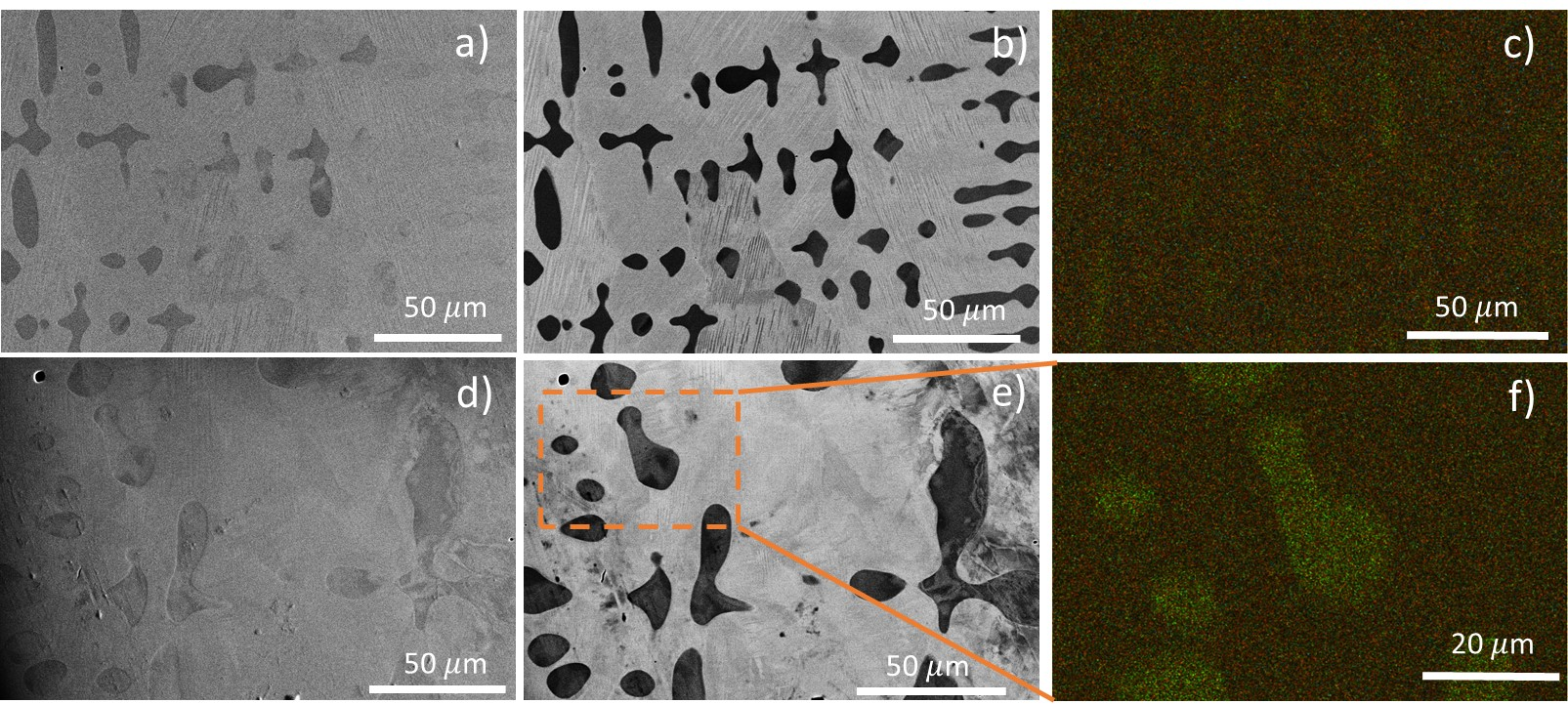}  
  \caption{SEM micrographs obtained using SE ---panels a) and d)--- and BSE ---panels b) and e)--- and corresponding compositional maps obtained by EDS ---panels c) and f)---. Panels a), b) and c) correspond to the ASB sample; and panels d), e) and f) refers to the TTB sample. Ni, Fe and Ga elements are represented in orange, green and blue, respectively. The dashed area in panel e) is magnified in panel f)}
  \label{fig:sem}
\end{figure*}

The microstructure and chemical composition of prepared samples have been analyzed by SEM. Figure~\ref{fig:sem} shows ---panels a) and d)--- the representative SEM secondary electron images of the polished samples. A small black dot in the panel d) corresponds to pores created during the fabrication process. The existence of gamma precipitates dispersed inside the martensite grains can be inferred from the SE images, which are clearly observed in the case of the backscattered images. The observation of the precipitates, which exhibit a larger size in the case of the TTB sample, is consistent with the XRD analysis reported above.

The EDS elemental mappings of the produced samples are depicted in Figure~\ref{fig:sem} ---panels c) and f)---, in which the signals of the \ce{Ni}, \ce{Fe} and \ce{Ga} elements are recorded. Table~\ref{tab:compAle} lists the composition of the martensite and $\gamma-$phases obtained by EDS point analysis of representative micro-areas of the samples. Despite the thermal treatment of the sample, which leads to the transformation from the non-modulated to the modulated martensitic structure, martensite phase maintains a stable composition. Accordingly, the same results have been obtained for the $\gamma-$phase, which is slightly enriched in \ce{Fe} and depleted in \ce{Ga} compared to the surrounding martensite matrix.

\begin{table*}
  \centering
  \begin{tabular}{lllllll}
    \toprule
    \textbf{Sample}&\multicolumn{3}{c}{\textbf{Martensite phase}}&\multicolumn{3}{c}{\textbf{$\mathbf{\gamma}-$phase}}\\
    \midrule
    &\ce{Ni}&\ce{Fe}&\ce{Ga}&\ce{Ni}&\ce{Fe}&\ce{Ga}\\
    \midrule
    ASB&\num{56\pm2}&\num{19\pm3}&\num{25\pm3}&\num{56\pm2}&\num{26\pm3}&\num{18\pm3}\\
    TTB&\num{56\pm2}&\num{19\pm3}&\num{25\pm3}&\num{57\pm2}&\num{27\pm3}&\num{16\pm3}\\
                                                                          \bottomrule
  \end{tabular}
  \caption{Chemical composition (atomic percentage) for the bulk samples obtained from EDS.}
  \label{tab:compAle}
\end{table*}

\subsection{DTA traces}
\label{sec:heat-flux}

The ASB sample with dimensions $\SI{9.0}{\milli\meter}\times\SI{9.57}{\milli\meter}\times\SI{2.35}{\milli\meter}$ ($\text{length}\times\text{width}\times\text{height}$) and $m=\SI{1.5987(1)}{\gram}$ in mass was placed in the conduction calorimeter to characterize its thermal properties. Thereafter the sample was removed from the calorimeter, quenched and transformed into the TTB sample, and was placed back in the calorimeter.

While inside the calorimeter the ASB was heated at \SI{335}{\kelvin}, in the austenite phase. Then the Julabo controller was set to \SI{300}{\kelvin} and the sample was cooled down against this fixed setpoint; the martensitic transformation took place. Thereafter, the sample ASB was further cooled down to \SI{270}{\kelvin} and heated back against a fixed setpoint of \SI{340}{\kelvin}, while the reverse transformation took place. A similar experiment was carried out for the TTB sample. The rate of temperature change in these runs was around \SI{1}{\kelvin\per\hour} (some \SI{0.015}{\kelvin\per\minute}), much slower than the rates in conventional DSC. We will refer this as a slow rate.

\begin{figure*}[!t]
  \centering
  \includegraphics[scale=0.85]{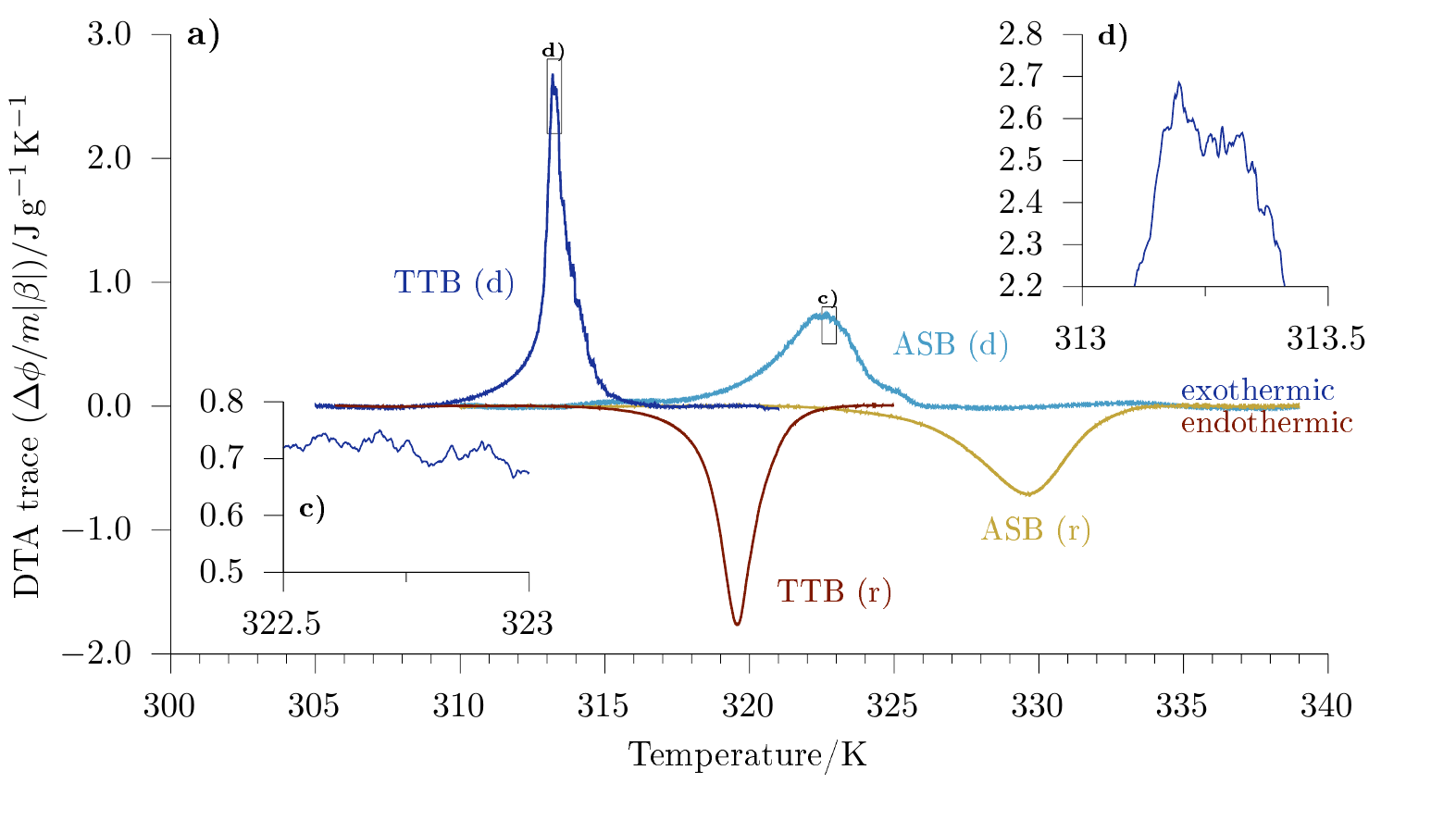}  

  \includegraphics[scale=0.85]{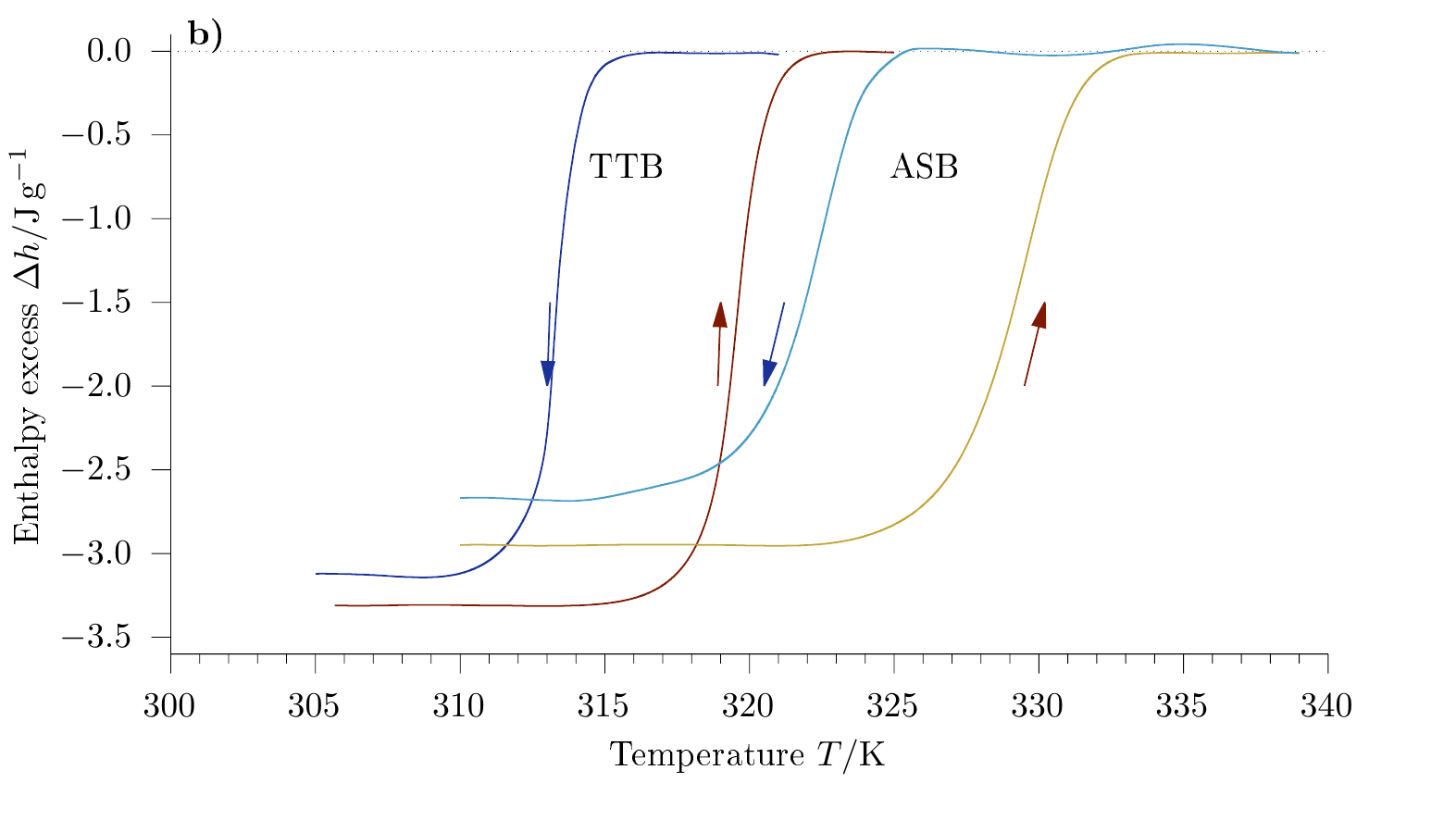}    
  \caption{The DTA traces (top, a) and the corresponding evolution of the enthalpy excesses (bottom, b) for the slow runs ($\beta\sim\SI{1}{\kelvin\per\hour}$) on a \usample{} bulk sample. The vertical axis displays excess heat per unit rate of temperature change and per unit mass (a) and excess enthalpy per unit mass (b). The ASB runs are shown in lighter shades; the TTB runs in darker shades pale. Blue is used for the direct, cooling, exothermic transformations; red for the reverse, heating, endothermic transformations. \textcolor{black}{Start, finish and peak temperatures are listed in Table~\ref{tab:temp}.} The insets zoom in the peak of the DTA anomaly to highlight the rough behaviour of the direct transformation in the ASB (c) and TTB (d) samples. The peak temperatures and the total enthalpy excesses are listed in Table~\ref{tab:temp}.}
  \label{fig:atd}
\end{figure*}

Figure~\ref{fig:atd}a shows the excesses of the DTA traces in a cooling (blueish) and a heating (reddish) runs.    The exothermic and endothermic excursions correspond to the direct (cooling) and reverse (heating) martensitic transformation.

\textcolor{black}{Table~\ref{tab:temp} lists the start and finish temperatures ---determined from the deviations of the DTA trace--- and the peak temperatures. The direct transformation  peaked at $M_p=\SI{\NTpeakd}{\kelvin}$, and the reverse transformation, at  $A_p=\SI{\NTpeakr}{\kelvin}$, for the ASB sample.}  On the other hand, the TTB showed cooler peak temperatures  $M_p=\SI{\Tpeakd}{\kelvin}$ and $A_p=\SI{\Tpeakr}{\kelvin}$, in agreement with Ref.~\cite{Manchon-Gordon2021}. The TTB sample also showed narrower anomalies and smaller thermal hysteresis $A_p-M_p=\SI{\hNT}{\kelvin}$ (ASB) vs $A_p-M_p=\SI{\hT}{\kelvin}$ (TTB). Thermal hysteresis was much smaller than previously reported values~\cite{Sarkar2014,Manchon-Gordon2021} ($\sim\SI{25}{\kelvin}$) for experiments carried out  at \num{1000} times faster rates of temperature change. 

Both cooling runs showed a distinct rough behaviour in the upper half of the direct transformation (see the insets I1 and I2 in Figure~\ref{fig:atd}), which was absent in the reverse transformation. This is evidence of a jerky behaviour in the direct transformation that will be later analyzed.

\begin{table*}
  \centering
\tabcolsep=10pt
\begin{tabular}{lS[table-format=3.1]S[table-format=3.1]S[table-format=3.1]S[table-format=3.1]S[table-format=3.1]S[table-format=3.1]S[table-format=3.1]S[table-format=3.1]}
\toprule
&\multicolumn{1}{c}{$M_s/\si{\kelvin}$}&\multicolumn{1}{c}{$M_p/\si{\kelvin}$}&\multicolumn{1}{c}{$M_f/\si{\kelvin}$}&\multicolumn{1}{c}{$A_s/\si{\kelvin}$}&\multicolumn{1}{c}{$A_p/\si{\kelvin}$}&\multicolumn{1}{c}{$A_f/\si{\kelvin}$}&\multicolumn{1}{c}{$\Delta h_d/\si{\joule\per\gram}$}&\multicolumn{1}{c}{$\Delta h_r/\si{\joule\per\gram}$}\\
\midrule
ASB&327&322.7&315&322&329.8&335&-2.7&-2.9\\
TTB&317.3&313.1&308.4&314.2&319.6&323.6&-3.1&-3.3\\
\bottomrule
\end{tabular}

  \caption{\textcolor{black}{Start ($s$), peak ($p$), and finish ($f$)  temperatures for the martensite (direct) transformation ($M$) and the austenite (reverse) transformation ($A$) in the \usample{} bulk sample without (ASB) and with (TTB) thermal treatment.} The transition enthalpies $\Delta h=h_{\text{low}}-h_{\text{high}}$ for the direct (cooling) and reverse (heating) transformation are also displayed. See Figure~\ref{fig:atd} for the DTA traces from which these data originated.}
  \label{tab:temp}
\end{table*}


The total enthalpy change ($\Delta h=h_{\text{low}}-h_{\text{high}}$) in either transformation is also listed in Table~\ref{tab:temp}. For the ASB sample we found $\Delta h_{\mathrm{d}}=\SI{\NTdhm}{\joule\per\gram}$ (direct) and $\Delta h_{\mathrm{r}}=\SI{\NTdha}{\joule\per\gram}$ (reverse). For the TTB sample, the excesses increased by \SI{20}{\percent} to $\Delta h_{\mathrm{d}}=\SI{\Tdhm}{\joule\per\gram}$ and $\Delta h_{\mathrm{r}}=\SI{\Tdha}{\joule\per\gram}$. These results are inline with those in Ref.~\cite{Sarkar2014} and Ref.~\cite{Manchon-Gordon2021} (see their Figure~7). The  evolution of the partial integration as a function of the temperature is shown in Figure~\ref{fig:atd}b.

\subsection{Avalanches}
\label{sec:avalanches}

The scanning rates of the previous experiments were much smaller than conventional DSC scans. The DTA cooling traces showed a distinct rough behavior which is magnified in the insets of Figure~\ref{fig:atd}. This is the result of the coalescence of a myriad of fast, microscopic exchanges, called jerks or avalanches that characterize the martensitic transformation. In our experiment, they occurred in a short scale of time compared to the scale of time of the experiment. Moreover, the insets of Figure~\ref{fig:atd} show 10 times larger excursions in the TTB sample. In Discussion we will develop arguments that link this behavior to the distribution of $\gamma-$phase in either sample.



The jerks or avalanches characterize, at the macroscopic scale, the energies associated with the intermittent dynamic of the phase transition as shown by  acoustic emission measurements~\cite{Vives1994}, calorimetry~\cite{Blobaum2006,Beke2018} and by high precision calorimetry~\cite{Gallardo2010,Vives2016,Romero2019,Romero2021a}. They are best analyzed in the adiabatic limit, when the thermal driving is slow enough to prevent overlapping. In that circumstances events are individualized, and their distribution is physically meaningful. Therefore, ultraslow experiments were conducted on the TTB sample following the strategy of~\citet{Romero2011}. Now the rate of temperature change was controlled by the Julabo FP45 and was set to $\beta_2=\SI{\pm40}{\milli\kelvin\per\hour}$, $\sim25$ slower than the previous run.


\begin{figure*}[t]
  \centering
  \includegraphics[width=\textwidth]{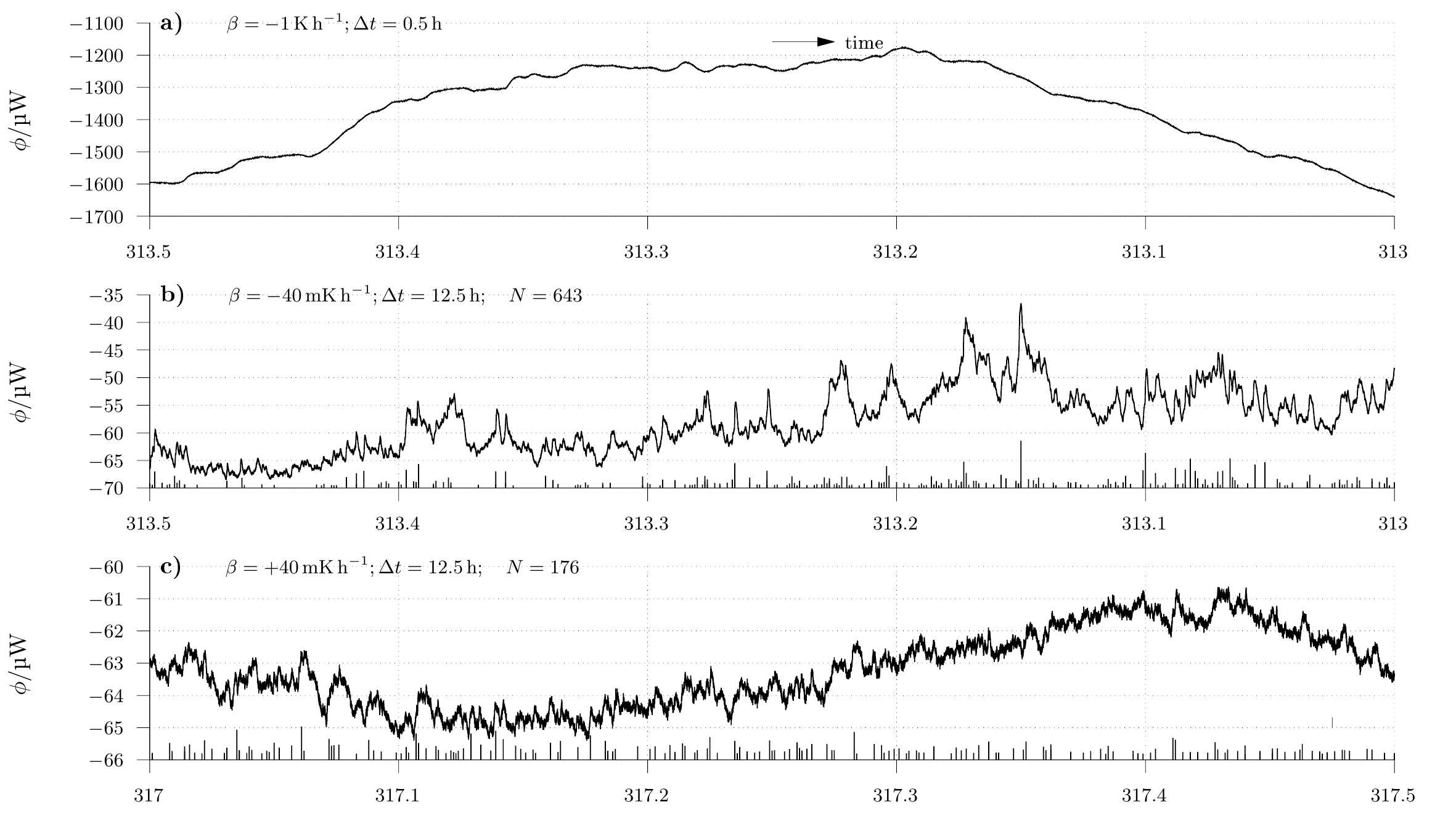}
    \caption{A close-up of the heat flux for $\beta=\SI{-1}{\kelvin\per\hour}$ (a, \SI{0.5}{\hour}),  $\beta=\SI{-40}{\milli\kelvin\per\hour}$ (b, \SI{12.5}{\hour}); and $\beta=\SI[retain-explicit-plus]{+40}{\milli\kelvin\per\hour}$ (c, \SI{12.5}{\hour}). Notice that the $x-$axis is conformed so that the time advances from left to right in every panel. Panel (b) and (c) show only one in every ten data points. The vertical axis in panel (c) is six times magnified relative to panel (b). The top left label in panels (b) and (c) display the number of jerks higher than \SI{30}{\nano\volt} in the plot. The jerks are located by vertical lines at the horizontal axis. The temperature interval shown in panel (b) and (c) are those of the highest activity in either run, see Figure~\ref{fig:bins}.} 
  \label{fig:picos}
\end{figure*}

Figure~\ref{fig:picos} shows an interval of \SI{0.5}{\kelvin} for the slow and the ultraslow runs. The panel a) shows the slow cooling run; panel b), the ultraslow cooling run; panel c), the ultraslow heating run. Notice that in panels a) and b) the temperature increases from right to left so that in every data set the time goes from left to right. Figure~\ref{fig:picos} shows the raw signal provided by the fluxmeters at a sampling rate of \SI{12.5}{\hertz} scaled by their sensitivity. In panel c) the standard noise of the signal ($\sim\SI{0.3}{\micro\watt}$ full height) is evident due to the high magnification of the vertical axis.


In order to characterize the spikes, we smoothed the cooling signal with a fifth-order all-pole Butterworth filter with a cut-off frequency of $\SI{1}{\hertz}$. On the other hand, the heating signal was resampled at \SI{0.1}{\hertz} by averaging blocks of \num{125} data points. Thereafter we identified local maxima and local minima: events when the signal stops increasing (maxima) or stops decreasing (minima). The size of the event (spike, jerk or avalanche) is associated with the size of a monotonous up rise from one minimum to the following maximum. The energy of the event is determined by scaling the heat flux with the time constant of the calorimeter ($\tau=\SI{70}{\second}$)~\cite{Gallardo2010}.

\begin{figure*}[t]
  \centering
  \includegraphics[width=\textwidth]{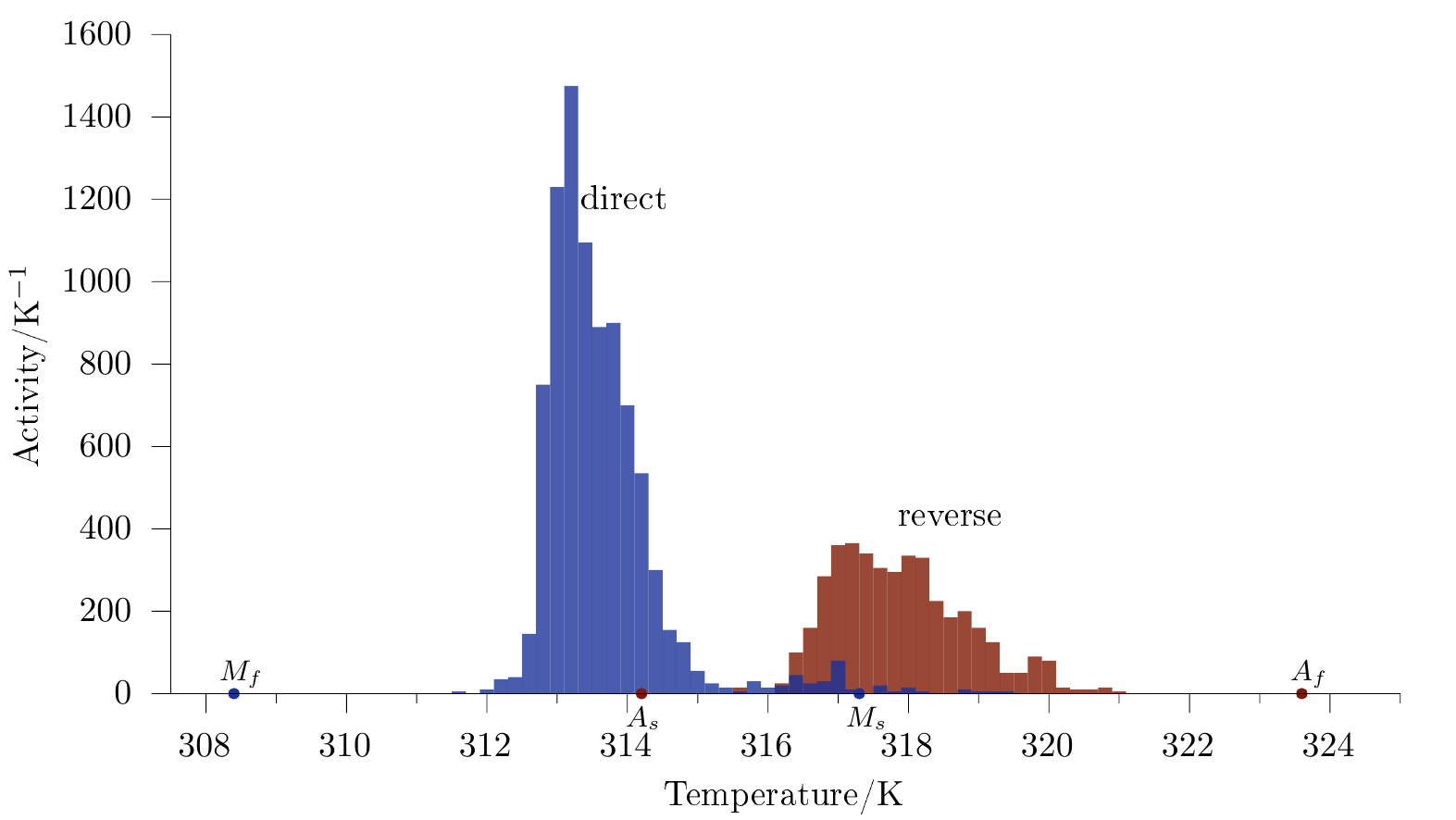}
  \caption{Histogram plot showing the activity ---number of events per kelvin--- derived from the number of events larger than $u_0=\SI{13}{\micro\joule}$ observed in a \SI{0.2}{\kelvin} interval for the ultraslow experiments. The bluish histogram shows the direct (cooling) run; the reddish histogram shows the reverse (heating) run. The $x-$axis shows the austenite and martensite start and finish temperatures, see Table~\ref{tab:temp}.}
  \label{fig:activity}
\end{figure*}

We searched for a threshold $E_0$ such that events larger or equal than $E_0$ occurred only during the temperature interval of the martensite transformation. Therefore, they can be attributed to the transformation. The threshold was identified at $E_0=\SI{30}{\nano\volt}$ in the raw signal of the voltmeter; $\phi_0=E_0/\alpha=\SI{0.2}{\micro\watt}$ in the heat flux scale and $u_0=\phi_0\tau=\SI{13}{\micro\joule}$ in the energetic scale. The top, left label in panels B and C of figure~\ref{fig:picos} shows the number of events larger than the threshold found in the interval.

 Figure~\ref{fig:activity} shows histogram plots of the activity ---number of events $E>E_0$ per kelvin--- measured along \SI{0.2}{\kelvin}-width bins for either ultraslow run. For the direct (cooling) transformation we collected $N=\num{1756}$ such events  with median value $\SI{26}{\micro\joule}$; they made $U=\SI{74}{\milli\joule}$, or \SI{1.5}{\percent} of the total enthalpy. For the reverse transformation we identified $N=\num{830}$ events, with median value $\SI{19}{\micro\joule}$ and making \SI{18}{\milli\joule} or \SI{0.4}{\percent} of the total enthalpy.

\newcommand{\eccdf}{\ensuremath{\mathrm{ECCDF}}}
\newcommand{\ccdf}{\ensuremath{\mathrm{CCDF}}}

\begin{figure*}
  \centering
  \includegraphics[width=\textwidth]{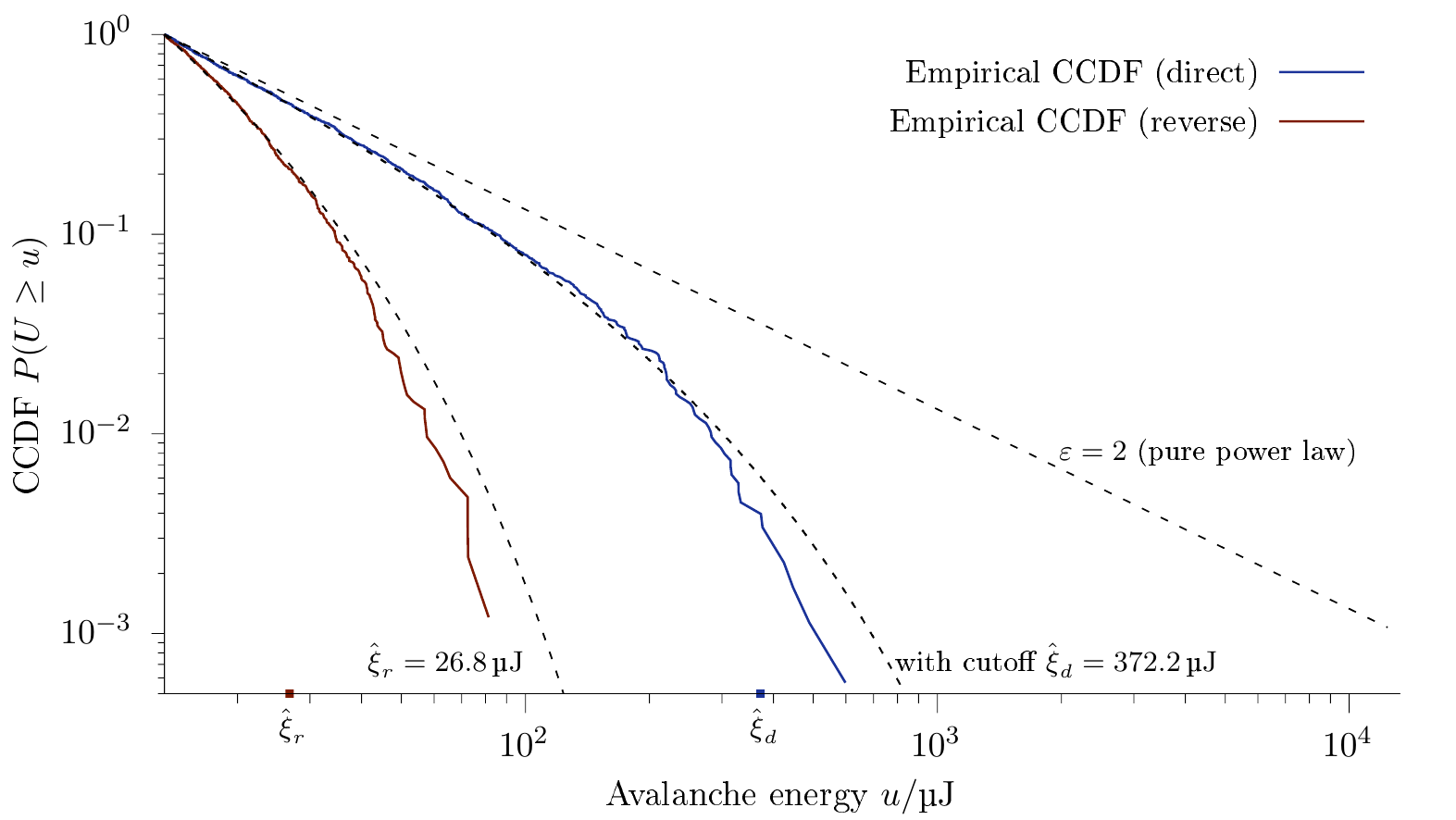}
  \caption{The energetic distributions of avalanche sizes for the ultraslow runs in the direct (dark blue) and reverse (dark red) transformations. The vertical axis displays the empirical complementary distribution function (ECCDF). The horizontal axis displays the energetic size of the avalanche. The quasilinear behaviour in the $\log-\log$ scale suggest a power law distribution with exponent $\varepsilon$ in the range $(2,3)$. The lack of very large energetic sizes suggest an exponential cutoff with characteristic energy $\hat\xi$.}
  \label{fig:bins}
\end{figure*}

Finally, we computed the empirical cumulative complementary distribution function (empirical \ccdf): the shares of avalanches larger or equal than a given size. The empirical \ccdf{} was determined by ranking the avalanches from the largest (rank one) to the smallest (rank $N$, assigned to $u_0$). Then, the empirical \ccdf{} is the ratio of the rank to the sample size.  Figure~\ref{fig:bins} shows  the empirical \ccdf{} against the avalanche size $u$ in the energetic scale for the direct (dark blue) and reverse (dark red) transformation. The empirical \ccdf{} shows a quasi linear behaviour in the $\log-\log$ plot, indicating a power law distribution of events $p(u)\propto u^{-\varepsilon}$, where the larger avalanches are increasingly scarce.

Notably, as observed in Figure~\ref{fig:picos} and in Figure~\ref{fig:atd}, the size of the largest avalanches during the reverse transformation is one order of magnitude smaller than those observed during the direct transformation.

These results show different behaviour in the direct and reverse transformations with greater activity, larger sizes and smaller exponent in the direct experiment. These differences  have already been reported and studied in martensitic transformations~\cite{Planes2017,Beke2018,Romero2019}.  In section~\ref{sec:discussion}, we will further discuss this issue.

\section{Discussion}
\label{sec:discussion}

Aside from the development of the 14M martensite phase, the main effect of thermal treatment is the coarsening of the $\gamma-$phase and the austenite crystals, which alter the differences between the direct and reverse transformations~\cite{Bolgar2017}. In Figure~\ref{fig:sem}, this is evidenced by the increase in the distances between $\gamma-$phase crystals.


The intermittent dynamics of the direct and the reverse transitions in martensitic transformations are characteristically asymmetric: exponents and accumulated energy differ in either transformations.  This has been previously associated with the fact that nucleation is required in the direct transformation, whereas the reverse transformation occurs by fast shrinkage of martensitic domains~\cite{Planes2017}. Alternatively, the different mechanism of the relaxation of elastic strain energy in either transition was proposed as a source of the asymmetry~\cite{Beke2018}. The asymmetry manifests itself even in the adiabatic conditions, see here Figure~\ref{fig:picos}, Figure~\ref{fig:activity}, Figure~\ref{fig:bins}, see also Ref.~\cite{Romero2019}, which put forward the intrinsic nature of the phenomenon. Analytically, the slope of the \ccdf{} at the top of the distribution characterizes the asymmetry. Our results in Figure~\ref{fig:bins} show $\varepsilon\sim2.3$ (direct) and  $\varepsilon\sim3$ (reverse). Either case the exponents are larger than $\varepsilon=2$, expected  for monoclinic to cubic transformation~\cite{Porta2019} and larger that the exponents reported previously on a \ce{NiFeGaCo} sample from DSC cooling measurements ($\varepsilon\sim\num{1.9}$), see Ref.~\cite{Bolgar2016,Bolgar2017}.

\citeauthor{Planes2017}~\cite{Planes2017} using acoustic emission characterized the asymmetry in, among others, \ce{Cu-Zn-Al} shape-memory alloy single crystals, which undergoes a cubic to monoclinic phase transition. They reported different exponents for the direct and reverse transformation and different distribution of energy sizes ---power law (direct) and power law with exponential cutoff (reverse)--- from a maximum likelihood analysis of the empirical observations. The analysis included a bidimensional chart of $\varepsilon,\xi$ for several choices of $E_{\min}$.

Inspired by this analysis, and considering that the grain boundaries must stop the growing of avalanches, we tested the distribution of calorimetric events against a power law with exponential cutoff for events larger than $u_0$ (see also Ref.~\cite{clauset-siam-2009}). The model is characterized by an exponent $\varepsilon$ and a damping energy $\xi$ which further hinders the probability of observing large events: $p(u;\varepsilon,\xi)\propto u^{-\varepsilon}\times\exp(-u/\xi)$. For our analysis we derived the empirical complementary cumulative distribution function \eccdf{} for the $N$ observed events larger than $u_0$ (see Section~\ref{sec:avalanches}). On the other hand, we set an array $200\times 200$ of $\varepsilon_i,\xi_j$ pairs and computed:
\begin{equation}
  \label{eq:2}
  \ccdf(u;\varepsilon_i,\xi_j)=\dfrac{\displaystyle\int\limits_u^\infty t^{-\varepsilon_i}\times\exp(-t/\xi_j)dt}{\displaystyle\int\limits_{u_0}^\infty t^{-\varepsilon_i}\times\exp(-t/\xi_j)dt},
\end{equation}
by numerical integration. The $\varepsilon$ values were arranged linearly from $1$ to $2.5$ (direct) and $1$ to $3.5$ (reverse); while the $\xi$ values were logarithmically arranged from $10^{0.5}u_0$ to $10^3u_0$.

Finally we computed the Kolmogorov-Smirnov distance given by the maximum absolute distance between the two, empirical and model, \ccdf{}:
\begin{equation}
  \label{eq:4}
  \begin{split}    
  &D(\varepsilon_i,\xi_j)=D_{ij}=\\&\sqrt{N}\times\max(|\eccdf(u)-\ccdf(u;\varepsilon_i,\xi_j)|).
\end{split}
\end{equation}
The Kolmogorov-Smirnov distance provides a $p-$value for the null hypothesis ``the empirical distribution originates from the tested model'': at the standard level of confidence $\alpha=\num{0.05}$ the null hypothesis sustains when $D$ is below $D_0=\num{1.358}$~\cite{Papoulis2002}. 

\begin{figure*}[t]
  \centering

  \includegraphics[width=\textwidth]{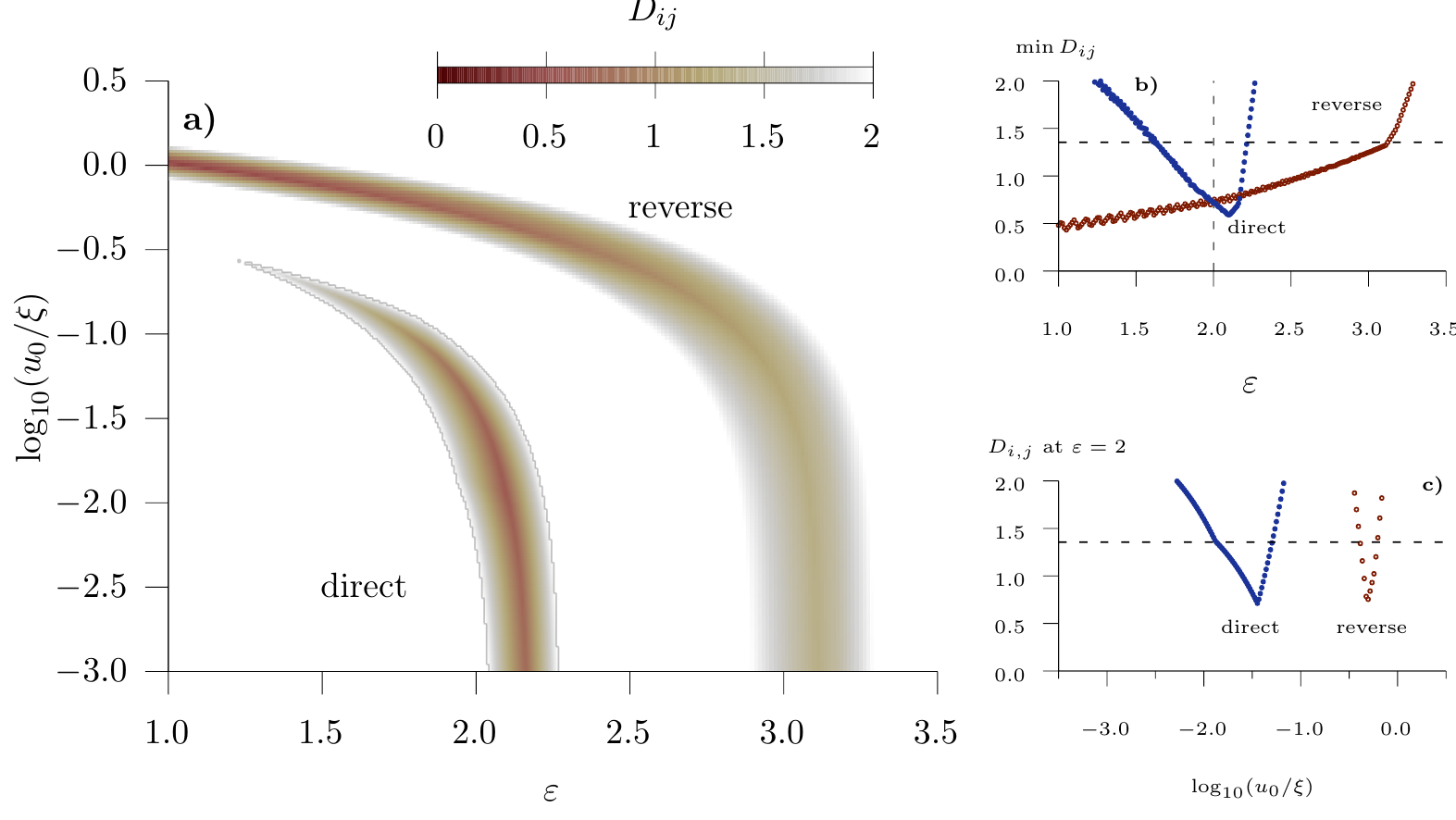}
  \caption{The Kolmogorov-Smirnov distance (color depth) from the empirical \ccdf{} to the \ccdf{} of a biparametric power-law distribution with exponential cutoff (a). The parameters of the model are represented in the $x-$axis (exponent) and $y-$axis (damping energy) in a $200\times200$ grid. The inset b) shows $\min(D_{ij})$ as a function of $\varepsilon$. The inset c) shows $D_{ij}$ as a function of $\xi$ for the condition $\varepsilon=2$, representative of the cubic to monoclinic transformation.  Either inset shows the direct transformation by solid blue circles and the reverse transformation by open red circles. Notice that either transformation attains similar, low values of $D_{ij}$ at $\varepsilon=2$ (b), whereas the damping parameter differs in one order of magnitude (c), see also Figure~\ref{fig:bins}.}
  \label{fig:ks}
\end{figure*}

 Figure~\ref{fig:ks}a shows the bi-dimensional grid $\varepsilon,\xi$ and, by the color depth, the Kolmogorov-Smirnov distance $D_{ij}$. The heatmap  highlights the region where $D_{ij}<2$: darker shades show smaller Kolmogorov-Smirnov distance and therefore greater similarity between the empirical distribution and the tested model. 

 The heatmap shows that for pure power law analysis ---that is $u_0/\xi\to0$--- the exponent goes to $\sim\num{2.3}$ (direct) and $\sim\num{3}$ (reverse), as noted earlier. Understandably, a finite damping energy yields a smaller exponent in the power law component. The inset b) shows the relationship between $\min(D_{ij})$ and $\varepsilon$ for the direct (solid blue circles) and the reverse (open red circles) transformations. The dashed line signals $D_0$ and, remarkably, both transformations sustains values below $D_0$ for $\varepsilon\sim2$. The inset c) shows the distribution of $D_{ij}$ as a function of $\xi$ at the condition $\varepsilon=2$. The local minimum $\hat\xi$ for either transformation can easily be spotted; they differ roughly in one order of magnitude.

 In summary a power-law with exponent $\varepsilon=2$ and an appropriate damping energy suffices to describe the distribution of events both in the direct and in the reverse transformation; the ratio $\Xi=\hat\xi_d/\hat\xi_r\sim\num{14}$ of the damping energies characterizes the asymmetry in the transformation. Figure~\ref{fig:bins} also shows the \ccdf{} for these choices of $\varepsilon=2,\hat\xi$ in either transformation (dashed lines), compared with the empirical distribution and the pure power-law behaviour.

 \begin{table*}
   \centering
   \begin{tabular}{lS[table-format=3.1]S[table-format=3.1]S[table-format=3.1]S[table-format=3.1]}
     \toprule
     &\multicolumn{4}{c}{$\log_{10}(u_0/\xi)$}\\
     \textbf{System}&\multicolumn{2}{c}{\textbf{AE}}&\multicolumn{2}{c}{\textbf{Calorimetry}}\\
                    &\multicolumn{1}{c}{Direct}&\multicolumn{1}{c}{Reverse}&\multicolumn{1}{c}{Direct}&\multicolumn{1}{c}{Reverse}\\
     \midrule
     \ce{Cu-Zn-Al}~\cite{Planes2017,Gallardo2010}&-2.0&\color{blue}-2.5&-1.3&-1.3\\
     \ce{Cu-Al-Be}~\cite{Romero2019}&-5.0&\color{blue}-2.0&-2.5&\color{blue}-1.0\\
     \ce{Ni-Fe-Ga-Co}~\cite{Bolgar2017}&-4.5&-3.0&-2.0&\multicolumn{1}{c}{n.a.}\\
     \ce{Ni-Mn-In}~\cite{Romero2021a}&-3.0&-3.0&-3.0&-3.0\\
     \midrule
     \ce{Ni-Fe-Ga} (this work)&&&\color{blue}-1.5&\color{blue}-0.3\\
     \bottomrule
   \end{tabular}
   \caption{Deduced upper bounduaries  (black) and reported values (blue) for $\log_{10}(u_0/\xi)$ in acoustic emission (AE) jerks and calorimetric jerks for system that undergo cubic to monoclinic martensitic transitions. The boundaries are given by $\log_{10}(u_0/u_{\max})$ when pure power-law is reported. It provides a lower bound since $\xi$ must be larger than $u_{\max}$. The rate of temperature change in AE experiments was usually in the range of several kelvin per minute. Calorimetric experiments were carried out in DSC calorimetry~\cite{Bolgar2017} at similar rates, and in conduction calorimetry (otherwise) at ultraslow rates.}
   \label{tab:previous}
 \end{table*}

Table~\ref{tab:previous} lists reported values from a set of previous AE and calorimetry studies. Pure power-law reports correspond to $\xi\to\infty$, but the study is always limited by the largest event in the catalog $u_{\max}$. Eventually the largest event is an experimental lower bound for $\xi$. The table reproduces this deduced lower bounds in black, and reported values of $\xi$ in blue ink. Values are shown as $\log_{10}(u_0/\xi)$.

In our case, we hypothesize that the arc melting technique followed by the thermal treatment has resulted in larger presence of local free energy barriers that lead to intermittent dynamics and asymmetry, as well as eventually prevented the growing of larger avalanches thus giving rise to a smaller, finite $\hat\xi$. Our estimate is some ten times smaller than results extracted from Ref.~\cite{Bolgar2017}.

On the other hand, if the transition mechanisms were associated with an elastic, surface energy~\cite{Beke2018} then $\sqrt{\Xi}\sim4$ would be a rough estimate of the ratio in the longitudinal size of the surface driving the avalanches. Alternatively, if they were associated with martensite phase developed in an austenite crystal~\cite{Planes2017}, $\sqrt[3]{\Xi}\sim2.5$ would be a rough estimate of the longitudinal size of the characteristic volume driving the avalanches. Furthermore, in this case  a rough estimation of an upper bound for the volume of sample affected in a single event can be obtained from the following argument.  We first take the total specific energy of the transformation ($|\Delta h|\sim\SI{3}{\joule\per\gram}$, see Table~\ref{tab:temp}), the sample mass $m$ and the sample volume $V$ (see Section~\ref{sec:experimental}). Then, the volume size of an event $e$ is proportional count $v=Ve/|\Delta h|m$ and the diameter of a sphere of equal volume is $d=6 (Ve/|\Delta h|m)^{1/3}/\pi$. For the median value of the event distribution ---$\langle e_d\rangle=\SI{26}{\micro\joule}$--- we get $d\sim\SI{100}{\micro\meter}$ which is roughly in agreement with the size of the austenite crystals that can be deduced from Figure~\ref{fig:sem} as this grain boundaries should stop the progression of martensite fronts during the transformation and would give rise to $\hat\xi_d=\SI{372.2}{\micro\joule}$, equivalent to the volume of a sphere \SI{300}{\micro\meter} in diameter. We note that less than \SI{1}{\percent} of the recorded events were larger than $\hat\xi_d$. In contrast, the reverse transition which occurs by fast shrinkage of the martensitic domains~\cite{Planes2017} showed a smaller damping energy $\hat\xi_r=\SI{25}{\micro\joule}$, similar to the median value of the event distribution ($\langle e_r\rangle=\SI{19}{\micro\joule}$) and resulted in roughly \SI{20}{\percent} of the recorded reverse events larger than $\xi_r$.



Accordingly, the lack of intermittent dynamics in the ASB samples suggests $\xi$ smaller than the observed values in the TTB sample. Understandably, the larger distribution of $\gamma-$phase should have further hindered the size of the events and resulted in a smoother distribution, see Figure~\ref{fig:atd}.



\section{Conclusions}
\label{sec:conclusion}

We have thermally characterized a \usample{} alloy produced by the arc melting technique with and without a posterior thermal treatment that promotes the metastable, adaptative 14M phase at room temperature.

The thermal treatment narrows the thermal hysteresis and the temperature distance for the start to finish condition. Interestingly, the thermal treatment enhances the intermittent dynamics (avalanches) of the martensite transition both in the direct and in the reverse transformations. The jerky behaviour was more prominent in the direct (cubic to 14M) transformation. 

The distribution of avalanches follows a power law with an exponential cutoff both in the direct and reverse transformation. The exponent $\varepsilon=2$ associated with the cubic to monoclinic transformation suffices to explain the power law decay both in the direct and reverse transformations. \textcolor{black}{This result is in line with previous observations}. \textcolor{black}{We identified the damping energies associated with the transformations and found that the reverse transformation is $10$ times more damped than the direct transformation. This is a quantification of the asymmetry of the processes involved in the transformation.}

\acknowledgments

A. Vidal-Crespo acknowledges a VI-PPITU fellowship from Universidad de Sevilla (Spain).

The authors wish to thank the staff at Departamento de F\'\i sica de la Materia Condensada and Facultad de Física for their continued support.

The X-ray fluorescence, the X-ray diffraction and SEM measurements were taken at the Centre for Research, Technology and Development (CITIUS) of the Universidad de Sevilla, with the financial support by Universidad de Sevilla.

The color squeme in Figure~\ref{fig:ks} is \texttt{bilbao} from Fabio Crameri's \emph{Scientific colour maps} \burl{https://www.fabiocrameri.ch/colourmaps/} \burl{https://doi.org/10.5281/zenodo.1243862}. Red and blue inks that represent the reverse and direct transformation come from Crameri's \texttt{roma} color squeme.


\section*{Declarations}

This work was supported by PAI of the Regional Government of Andalusia, VI PPITU and VII PPITU of Universidad de Sevilla, and  by Junta de Andaluc\'\i a-Consejer\'\i a de Conocimiento, Investigaci\'on y Universidad project ProyExcel\_00360.

The authors declare no competing interests.

Data are available upon reasoned request to the corresponding author. 

\section*{CRediT authorship contribution statement}
\label{sec:credi}

Conceptualization: Francisco Javier Romero; Javier S. Bl\'azquez.

Methodology: Jos\'e-Mar\'\i a Mart\'\i n-Olalla; Francisco Javier Romero.

Data Curation: Jos\'e-Mar\'\i a Mart\'\i n-Olalla; Antonio Vidal-Crespo.

Formal Analysis and Investigation: Jos\'e-Mar\'\i a Mart\'\i n-Olalla; Antonio Vidal-Crespo; Francisco Javier Romero; Alejandro F. Manch\'on-Gord\'on; Jhon J. Ipus; Mar\'\i a Carmen Gallardo.

Original draft preparation: Jos\'e Mar\'\i a Mart\'\i n-Olalla; Alejandro F. Manch\'on-Gord\'on; Javier S. Bl\'azquez.

Review and editing: Antonio Vidal-Crespo; Francisco J Romero; Jhon J. Ipus; María Carmen Gallardo; Clara F Conde.

Supervision: Maria Carmen Gallardo; Clara F Conde.

%



\end{document}